%% file: paper.tex
\documentclass[preprint2, times, twocolappendix]{aastex631}

\input{tex/build/macros/project.tex}
\input{tex/src/preamble-maneage.tex}

\input{tex/src/preamble-project.tex}

\begin{document}

%% Title
\title{\projecttitle}

\author[0000-0002-6672-1199]{Sepideh Eskandarlou}
\affiliation{Centro de Estudios de F\'isica del Cosmos de Arag\'on (CEFCA), Plaza San Juan 1, 44001 Teruel, Spain}

\correspondingauthor{Sepideh Eskandarlou}
\email{sepideh.eskandarlou@gmail.com}

\author[0000-0003-1710-6613]{Mohammad Akhlaghi}
\affiliation{Centro de Estudios de F\'isica del Cosmos de Arag\'on (CEFCA), Plaza San Juan 1, 44001 Teruel, Spain}

\author[0000-0002-6220-7133]{Ra\'ul Infante-sainz}
\affiliation{Centro de Estudios de F\'isica del Cosmos de Arag\'on (CEFCA), Plaza San Juan 1, 44001 Teruel, Spain}

\author[0000-0002-5075-1764]{Elham Saremi}
\affiliation{Instituto de Astrof\'isica de Canarias, Calle V\'ia L\'actea s/n, 38205 La Laguna, Spain}
\affiliation{Departamento de Astrof\'isica, Universidad de La Laguna, 38205 La Laguna, Spain}
\affiliation{School of Physics \& Astronomy, University of Southampton, Highfield Campus, Southampton SO17 1BJ, UK}

\author[0000-0001-9000-5507]{Samane Raji}
\affiliation{Dept. of Theoretical and Atomic Physics, and Optics, University of Valladolid, Spain}

\author[0009-0004-5054-5946]{Zahra Sharbaf}
\affiliation{Instituto de Astrof\'isica de Canarias, Calle V\'ia L\'actea s/n, 38205 La Laguna, Spain}
\affiliation{Departamento de Astrof\'isica, Universidad de La Laguna, 38205 La Laguna, Spain}

\author[0009-0001-2377-272X]{Giulia Golini}
\affiliation{Instituto de Astrof\'isica de Canarias, Calle V\'ia L\'actea s/n, 38205 La Laguna, Spain}
\affiliation{Departamento de Astrof\'isica, Universidad de La Laguna, 38205 La Laguna, Spain}

\author[0000-0002-6467-8078]{Zohreh Ghaffari}
\affiliation{Instituto de Astrof\'isica de Canarias, Calle V\'ia L\'actea s/n, 38205 La Laguna, Spain}
\affiliation{Departamento de Astrof\'isica, Universidad de La Laguna, 38205 La Laguna, Spain}
\affiliation{INAF-Osservatorio Astronomico di Trieste, via G.B. Tiepolo, 11-I-34143 Trieste, Italy}

\author[0000-0003-1643-0024]{Johan H. Knapen}
\affiliation{Instituto de Astrof\'isica de Canarias, Calle V\'ia L\'actea s/n, 38205 La Laguna, Spain}
\affiliation{Departamento de Astrof\'isica, Universidad de La Laguna, 38205 La Laguna, Spain}

%% Abstract
\begin{abstract}
  \noindent
Calibration of pixel values is a fundamental step for accurate measurements in astronomical imaging.
In astronomical jargon this is known as estimating zero point magnitude.
Here, we introduce a newly added script in GNU Astronomy Utilities (Gnuastro) version 0.20 for the zero point magnitude estimation, named: \texttt{astscript-zeropoint}.
The script offers numerous features, such as the flexibility to use either image(s) or a catalog as the reference dataset.
Additionally, steps are parallelized to enhance efficiency for big data.
Thanks to Gnuastro's minimal dependencies, the script is both flexible and portable.
The figures of this research note are reproducible with Maneage, on the Git commit \projectversion.
\end{abstract}

%% Keywords (from https://astrothesaurus.org)

%% Start of main body.
\section{Introduction}\label{sec:intro}
\noindent
Scientific measurements of astronomical sources are not possible without reduction and calibration of the raw data.
In astronomical imaging, the calibration of pixel values is done through the \emph{zero point} magnitude which allows the conversion of the pixel value units to physical (for example SI or CGS) units.
It encodes all the instrument-specific and observational factors into a single number.
Therefore, after accounting for the zero point, measurements like flux, brightness, and other parameters can be compared with other measurements from different telescopes and instruments.
This allows the use of the data for higher-level analysis, for instance, measuring the galaxy's stellar masses or spectral energy distribution fitting.

In this research note, we introduce a new script in charge of computing the zero point magnitude of astronomical images with the executable name of `\texttt{astscript-zeropoint}'.
It performs this by aperture photometry of stars in the image and comparing them with already calibrated datasets: image(s) or a catalog.
This script was introduced into GNU Astronomy Utilities, or Gnuastro\footnote{\url{https://www.gnu.org/software/gnuastro}} \citep{gnuastro,gnuastro2019} since version 0.20.
Earlier versions of this script have been used by other research projects such as \citet{trujillo2021} and \citet{martinez2021}.

\section{Computing the zero point magnitude}\label{sec:computezp}

The `\texttt{astscript-zeropoint}' script requires reference image(s) or a catalog that should be calibrated and overlap with the image on the sky.
This reference data is used as the basis to compare the aperture photometry conducted by the script.
The basic steps performed internally by the script are:

\begin{itemize}
  \setlength\itemsep{-1mm}
 \item Query the Gaia \citep{gaia2016} catalog to obtain accurate coordinates of stars in an input image.
 \item Perform aperture photometry considering different aperture sizes.
 \item For each aperture and each star, compute the difference in magnitude with respect to the reference magnitude of that star ($\delta_{a,s}$).
Setting the input zero point magnitude to zero.
\item Compute the zero point magnitude for each aperture ($z_a$) from the sigma-clipped median of the $\delta_{a,s}$ distribution for all stars.
  This can be either from the full sample, or optionally, in a specified magnitude range (to avoid the noisy faint stars and bright saturated stars).
 \item The final zero point magnitude is the $z_a$ with smallest scatter (sigma-clipped standard deviation, $\sigma$, of the $\delta_{a,s}$ distribution).
\end{itemize}

Depending on the reference data (images or a catalog), the execution of the script is slightly different: the former requires generation of a catalog through aperture photometry.
A tutorial on the usage of both scenarios is available in the Tutorials chapter\footnote{\url{https://www.gnu.org/software/gnuastro/manual/html_node/Tutorials.html}} of the manual.
The tutorial includes examples in both scenarios (of using references images or a catalog).

\begin{figure*}[!t]
  \begin{center}
  \ifdefined\makepdf%
    \tikzsetnextfilename{fig-zp}%
    \input{tex/src/fig-zp.tex}%
  \else
    \includegraphics[width=\linewidth]{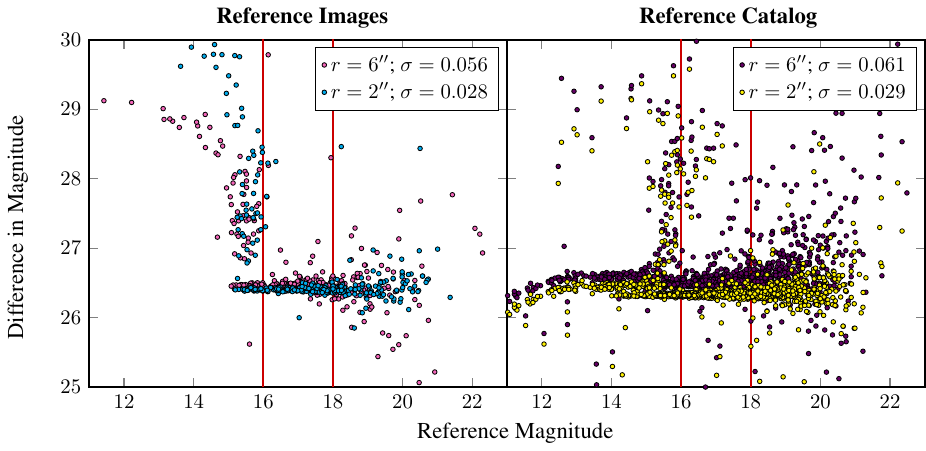}
  \fi

  \caption{\label{fig:refimg} Difference in magnitude (using apertures of radius {\figapermin}$^{\prime\prime}$ and {\figapermax}$^{\prime\prime}$) as a function of the reference magnitude for stars in the input image.
    The horizontal range of this plot shows the zero point magnitude and its scatter.
    In both cases SDSS DR12 is used as a reference: the left panel uses an SDSS image and right uses on the SDSS catalog overlapping with our input image.
    In right panel (from a catalog), stars brighter than the 15th magnitude are present because the SDSS pipeline attempts to find the magnitude of saturated stars (which is not done in the left panel).
    The blue and yellow filled circles have a smaller dispersion ($\sigma$ in legend).
    The zero point magnitudes are obtained based on the horizontal component of the blue and yellow filled circles with lower $\sigma$ (within the magnitude range that is shown with the red vertical lines).
    The underlying data for creating the left and right panels are available in \href{http://doi.org/10.5281/zenodo.10256845}{zenodo:10256845} (files with the \texttt{.txt} suffix, not changed since Commit e31dd15).}
  \end{center}
\end{figure*}

Figure \ref{fig:refimg} shows the difference in magnitude of apertures as a function of the reference magnitude (calculated from apertures in images and taken directly from the catalog) with lowest and highest $\sigma$.
The input image is from J-PLUS \citep{cenarro19}.
The ultimate zero point magnitude comes from the aperture with least scatter in the magnitude range between {\figmagmin} to {\figmagmax} for this particular example.
For instance in the examples of Figure \ref{fig:refimg} the apertures with radius {\figbestaper} arc-seconds have smallest $\sigma$ with zero point of {\zptwoimg} and {\zptwocat} using reference images and catalog respectively.
The break at brighter magnitudes (between magnitudes {\figmagsplittwominsdss} to {\figmagsplittwomaxsdss}) that we see in the reference catalog plot is created because SDSS models the central parts of saturated stars in its catalog.
More details and complete information on all run-time options of this script can be found under the ``Invoking astscript-zeropoint'' section\footnote{\url{https://www.gnu.org/software/gnuastro/manual/html\_node/Invoking-astscript\_002dzeropoint.html}} of Gnuastro's manual.

\section{Acknowledgement}
We gratefully acknowledge Ignacio Trujillo for discussions that improved the script.

The workflow of this research note was developed in the reproducible framework of Maneage \citep[\emph{Man}aging data lin\emph{eage},][latest Maneage commit \maneageversion{}, from \maneagedate]{maneage}.
This note is created from the Git commit {\projectversion} that is hosted on Codeberg\footnote{In the \texttt{zeropoint} branch of \url{\projectgitrepo}} and is archived on SoftwareHeritage\footnote{\href{https://archive.softwareheritage.org/swh:1:dir:8b2d1f63be96de3de03aa3e2bb68fa7fa52df56f;origin=https://codeberg.org/gnuastro/papers;visit=swh:1:snp:e37e226bab517eef24d854467682b2fcf5d7dc32;anchor=swh:1:rev:ea682783d83707c0e1d114a5de74a100be9f545d}{swh:1:dir:8b2d1f63be96de3de03aa3e2bb68fa7fa52df56f} Software Heritage identifiers (SWHIDs) can be used with resolvers like \texttt{http://n2t.net/} (e.g., \texttt{http://n2t.net/swh:1:...}). Clicking on the SWHIDs will provide more “context” for same content} for longevity.

The analysis of this research note was done using Gnuastro (ascl.net/1801.009) version \gnuastroversion. Work on Gnuastro has been funded by the Japanese Ministry MEXT scholarship and its Grant-in-Aids 21244012 and 24253003, ERC 339659-MUSICOS, the Spanish grants AYA2016-76219-P and PID2021-124918NA-C43 as well as the NextGenerationEU ICTS-MRR-2021-03-CEFCA.

%% Bibliography
\bibliography{references}{}
\bibliographystyle{aasjournal}

%% Start appendix.
\appendix

%% Mention all used software in an appendix.
\section{Software acknowledgement}
\label{appendix:software}
\input{tex/build/macros/dependencies.tex}

%% Finish LaTeX
\end{document}

%% file: tex/build/macros/project.tex
\input{tex/build/macros/initialize.tex}
\input{tex/build/macros/fig-zp.tex}
\input{tex/build/macros/verify.tex}
\input{tex/build/macros/hardware-parameters.tex}

%% file: tex/build/macros/initialize.tex
\newcommand{\projecttitle}{Gnuastro: Estimating the Zero Point Magnitude in Astronomical Imaging}
\newcommand{\projectversion}{c89275e}
\newcommand{\projectgitrepo}{https://codeberg.org/gnuastro/papers}
\newcommand{\projectcopyrightowner}{Sepideh Eskandarlou <sepideh.eskandarlou@gmail.com>}
\newcommand{\gnuastroversion}{0.21}
\newcommand{\maneagedate}{22 May 2023}
\newcommand{\maneageversion}{8161194}

%% file: tex/build/macros/fig-zp.tex
\newcommand{\zptwoimg}{26.405}
\newcommand{\zptwoimgstd}{0.028}

\newcommand{\zpsiximgstd}{0.056}
\newcommand{\zptwocat}{26.334}
\newcommand{\zptwocatstd}{0.029}

\newcommand{\zpsixcatstd}{0.061}
\newcommand{\figmagmin}{16}
\newcommand{\figmagmax}{18}
\newcommand{\figapermin}{2}
\newcommand{\figapermax}{6}
\newcommand{\figbestaper}{two}
\newcommand{\figmagsplittwominsdss}{14}
\newcommand{\figmagsplittwomaxsdss}{16}

%% file: tex/src/preamble-maneage.tex
\ifdefined\highlightnew

\else

\fi

\ifdefined\highlightnotes
\newcommand{\tonote}[1]{\textcolor{red!60!black}{[#1]}}
\else
\newcommand{\tonote}[1]{{}}
\fi

%% file: tex/src/preamble-project.tex
\usepackage{graphicx}

\usepackage{xcolor}
\color{black} % Color of main text.
\definecolor{DarkBlue}{RGB}{0,0,90}

\hypersetup{
    pdftitle={\projecttitle},
    pdfauthor={\projectcopyrightowner},
    pdfsubject={\projectgitrepo{} (commit \projectversion)},
    pdfkeywords={Reproducible research, Maneage, ADD YOUR OWN}
}

\input{tex/src/preamble-pgfplots.tex}

\shorttitle{\projecttitle}
\shortauthors{Eskandarlou et al.}

%% file: tex/src/preamble-pgfplots.tex
\usepackage{tikz}
\usetikzlibrary{external}
\tikzsetexternalprefix{tex/tikz/}

\usepackage{pgfplots}
\pgfplotsset{compat=newest}
\usepgfplotslibrary{groupplots}
\pgfplotsset{
  axis line style={thick},
  tick style={semithick},
  tick label style = {font=\footnotesize},
  every axis label = {font=\footnotesize},
  legend style = {font=\footnotesize},
  label style = {font=\footnotesize}
  }

%% file: tex/src/fig-zp.tex
%% Insert the images.
\begin{tikzpicture}

  \begin{groupplot}[
      group style={group size=2 by 1,
                   horizontal sep=0pt,
                   vertical sep=0pt,
                   xticklabels at=edge bottom,
                   yticklabels at=edge left},
      grid=minor,
      xmin=11,
      xmax=23,
      ymin=25,
      ymax=30,
      width=0.48\linewidth,
      enlarge x limits=false,
      enlarge y limits=false,
      ]

    \nextgroupplot

    %% Add radial profile.
    \addplot [only marks, fill=magenta!60!white, mark size=1.1]
    table {tex/build/figures/zp-with-ref-img/6-arcsec-zp-with-ref-img.txt};
    \addlegendentry{$r={\figapermax}^{\prime\prime}$; $\sigma={\zpsiximgstd}$}

    %% Add radial profile.
    \addplot [only marks, fill=cyan, mark size=1.1]
    table {tex/build/figures/zp-with-ref-img/2-arcsec-zp-with-ref-img.txt};
    \addlegendentry{$r={\figapermin}^{\prime\prime}$; $\sigma={\zptwoimgstd}$}

    %% Add line to show in which magnitude range the stars are selected
    \addplot[mark=none, red!80!black, line width=1pt] coordinates {({\figmagmin},25) ({\figmagmin},32)};
    \addplot[mark=none, red!80!black, line width=1pt] coordinates {({\figmagmax},25) ({\figmagmax},32)};

    \nextgroupplot

   %% Add radial profile.
    \addplot [only marks, fill=violet!80!black, mark size=1.1]
    table {tex/build/figures/zp-with-ref-cat/6-arcsec-zp-with-ref-cat.txt};
    \addlegendentry{$r={\figapermax}^{\prime\prime}$; $\sigma={\zpsixcatstd}$}

    %% Add radial profile.
    \addplot [only marks, fill=yellow, mark size=1.1]
    table {tex/build/figures/zp-with-ref-cat/2-arcsec-zp-with-ref-cat.txt};
    \addlegendentry{$r={\figapermin}^{\prime\prime}$; $\sigma={\zptwocatstd}$}

    %% Add line to show in which magnitude range the stars are selected
    \addplot[mark=none, red!80!black, line width=1pt] coordinates {({\figmagmin},25) ({\figmagmin},32)};
    \addplot[mark=none, red!80!black, line width=1pt] coordinates {({\figmagmax},25) ({\figmagmax},32)};

  \end{groupplot}

  \node[anchor=south] at (0.4\linewidth,-0.06\linewidth)
       {Reference Magnitude};

  \node[anchor=south, rotate=90] at (-0.05\linewidth,0.15\linewidth)
       {Difference in Magnitude};

  \node[anchor=south] at (0.6\linewidth,0.33\linewidth)
       {\bf Reference Catalog};

  \node[anchor=south] at (0.2\linewidth,0.33\linewidth)
       {\bf Reference Images};

\end{tikzpicture}

%% file: tex/build/macros/dependencies.tex
 
This research was done with the following free software programs and libraries:  1.23, Bzip2 1.0.8, CFITSIO 4.1.0, CMake 3.24.0, cURL 7.84.0, Dash 0.5.11-057cd65, Discoteq flock 0.4.0, Expat 2.4.1, File 5.42, Fontconfig 2.14.0, FreeType 2.11.0, Git 2.37.1, GNU Astronomy Utilities 0.21 \citep{gnuastro,gnuastro2019}, GNU Autoconf 2.71, GNU Automake 1.16.5, GNU AWK 5.1.1, GNU Bash 5.2-rc2, GNU Binutils 2.39, GNU Bison 3.8.2, GNU Compiler Collection (GCC) 12.1.0, GNU Coreutils 9.1, GNU Diffutils 3.8, GNU Findutils 4.9.0, GNU gettext 0.21, GNU gperf 3.1, GNU Grep 3.7, GNU Gzip 1.12, GNU Integer Set Library 0.24, GNU libiconv 1.17, GNU Libtool 2.4.7, GNU libunistring 1.0, GNU M4 1.4.19, GNU Make 4.3, GNU Multiple Precision Arithmetic Library 6.2.1, GNU Multiple Precision Complex library, GNU Multiple Precision Floating-Point Reliably 4.1.0, GNU Nano 6.4, GNU NCURSES 6.3, GNU Readline 8.2-rc2, GNU Scientific Library 2.7, GNU Sed 4.8, GNU Tar 1.34, GNU Texinfo 6.8, GNU Wget 1.21.2, GNU Which 2.21, GPL Ghostscript 9.56.1, Help2man , Less 590, Libffi 3.4.2, Libgit2 1.3.0, libICE 1.0.10, Libidn 1.38, Libjpeg 9e, Libpaper 1.1.28, Libpng 1.6.37, libpthread-stubs (Xorg) 0.4, libSM 1.2.3, Libtiff 4.4.0, libXau (Xorg) 1.0.9, libxcb (Xorg) 1.15, libXdmcp (Xorg) 1.1.3, libXext 1.3.4, Libxml2 2.9.12, libXt 1.2.1, Lzip 1.23, OpenSSL 3.0.5, PatchELF 0.13, Perl 5.36.0, pkg-config 0.29.2, podlators 4.14, Python 3.10.6, util-Linux 2.38.1, util-macros (Xorg) 1.19.3, WCSLIB 7.11, X11 library 1.8, XCB-proto (Xorg) 1.15, xorgproto 2022.1, xtrans (Xorg) 1.4.0, XZ Utils 5.2.5 and Zlib 1.2.11. 
The \LaTeX{} source of the paper was compiled to make the PDF using the following packages: courier 61719 (revision), epsf 2.7.4, etoolbox 2.5k, helvetic 61719 (revision), lineno 5.3, pgf 3.1.10, pgfplots 1.18.1, revtex4-1 4.1s, tex 3.141592653, textcase 1.04 and ulem 53365 (revision). 
We are very grateful to all their creators for freely  providing this necessary infrastructure. This research  (and many other projects) would not be possible without  them.